\documentclass[pre,twocolumn,amsmath,amssymb,floatfix]{revtex4-1}

\usepackage{graphicx}
\usepackage{dcolumn}
\usepackage{bm}
\usepackage{color}
\usepackage{mciteplus}
\usepackage{braket}

\begin{document}

\title{Absorption of waves by large scale winds in stratified turbulence}
\author{P.~Clark di Leoni, P.D.~Mininni}
\affiliation{Departamento de F\'\i sica, Facultad de Ciencias 
Exactas y Naturales, Universidad de Buenos Aires and IFIBA, CONICET, 
Ciudad Universitaria, 1428 Buenos Aires, Argentina.}
\date{\today}

\begin{abstract}
The atmosphere is a nonlinear stratified fluid in which internal 
gravity waves are present. These waves interact with the flow, resulting
in wave turbulence that displays important differences with the
turbulence observed in isotropic and homogeneous flows. We study
numerically the role of these waves and their interaction with the large
scale flow, consisting of vertically sheared horizontal winds. We
calculate their space and time resolved energy spectrum (a 
four-dimensional spectrum), and show that most of the energy
is concentrated along a dispersion relation that is Doppler shifted by
the horizontal winds. We also observe that when uniform winds are let
to develop in each horizontal layer of the flow, waves whose phase
velocity is equal to the horizontal wind speed have negligible energy.
This indicates a nonlocal transfer of their energy to the mean flow.
Both phenomena, the Doppler shift and the absorption of waves traveling
with the wind speed, are not accounted for in current theories of
stratified wave turbulence.
\end{abstract}

\maketitle

\section{Introduction} 

Turbulent flows are often pictured as highly disorganized, with energy being
transferred from large scale motions to small scale eddies. But in some cases
the opposite can happen. In 1967 Kraichnan \cite{Kraichnan80} developed the
theory of the inverse energy cascade, in which nonlinearities transfer energy
towards larger structures in a self-organization process. The theory had a huge
impact in oceanography and meteorology, and in the basic physical understanding
of turbulence. However, an inverse cascade is not the only possible
mechanism by which energy can be transferred from small to large
scales, and in some flows other mechanisms can result in the
generation of large scale flows.

An important example is given by stratified flows. Stratification
plays a key role in the dynamics of the oceans and the
atmosphere. As an example, internal gravity waves arising from it are
responsible for transfer of momentum between different regions of the
atmosphere \cite{Hines72}, and for transport in the oceans
\cite{Ledwell00,Gargett04}.  Much effort has been put in characterizing
internal gravity waves and their relation with geophysical flows
\cite{Riley00,Staquet02,Ivey08}.

Several observational studies suggest there is a coupling between the
mean (or large-scale) flow and internal gravity waves
\cite{Finnigan84,Dohan03,Xing05}.  As a result, various models for the
interaction between a wave field and a mean background flow have been
put forward \cite{Grimshaw72,Grimshaw75,Hines91}.  These models, which
were mainly developed in the context of atmospheric sciences and
oceanography, predict that the dispersion relation of the waves is
Doppler shifted by the background flow.  Furthermore, they also describe
an instability by which waves whose horizontal phase velocity matches
that of the horizontal wind speed on a given layer of the fluid, are
absorbed by the background flow.  This is the so-called critical layer
(CL) instability. A fraction of the energy in the waves is then transferred 
towards the mean flow, and the rest is dissipated. However, most CL 
models are either linear or perturbative, have the large scale shear 
externally imposed, lack nonlinear coupling between the waves 
themselves as well as with small-scale eddies, and cannot fully 
describe the nonlinear dynamics of a turbulent geophysical flow. 
Important observational studies of the atmosphere \cite{Gossard70} 
and the ocean \cite{Kunze90}, and laboratory experiments 
\cite{Thorpe81,Koop86} focusing on the properties of single waves 
passing through, show Doppler shift and indications of CL instability, 
although they lack the presence of a fully developed turbulent 
superposition of waves. In the same spirit, numerical studies of 
Doppler shift and CL instability 
\cite{Grimshaw75,Hines91,Winters94,Broutman97} that solve nonlinear
atmospheric models, focus on single wave packets interacting with a
background flow which is externally imposed, and do not take into
account the turbulent nature of geophysical flows nor the nonlinear
interaction between the internal gravity waves themselves and with the
eddies present in the flow (see however \cite{Winters94}, where the 
full nonlinear equations are solved, although a single wavepacket
and background flow are imposed).  This can be understood as
identifying waves in a smooth flow can be done by observing the time
evolution of the system, while extraction of the waves in a disordered
background requires space and time information.

On the other hand, fluid dynamics studies, both theoretically and
numerically (see, e.g.,
\cite{Billant01,Waite06,Brethouwer07,Lindborg07}), focused mainly on the
fully nonlinear aspect of stratified turbulence, on how energy is
distributed among scales, and on the development of anisotropy and of
flat ``pancake'' structures in the flow. These studies remark that the
interaction between waves and eddies is of fundamental importance in
stratified turbulence, but have neglected the effects of Doppler
shifts or CL instability. An important topic which is subject to an
active debate is the existence of an inverse cascade in purely
stratified flows. As already mentioned, an inverse cascade is a 
self-similar process by which energy is transferred nonlinearly from 
small scales to larger scales, so energy gets accumulated in
structures with larger correlation length than that of the injection
 mechanism. Inverse cascades have been extensively studied in 
two dimensional turbulence \cite{Kraichnan80}, and have also been 
observed in rotating turbulence (including rotating stratified 
turbulence) \cite{Bartello95,Pouquet13,Yarom14}. But results for 
purely stratified turbulence are unclear. It is known that stratified 
flows can generate large-scale vertically sheared horizontal
winds (VSHW) \cite{Smith02}, although the mechanism involved is not
entirely understood.  While evidence of energy flow toward large scales
has been reported \cite{Marino13,Marino14}, several authors have argued
that an inverse cascade is not possible \cite{Waite06,Herbert14} based
on statistical mechanical arguments.

Recently, weak turbulence theory \cite{Nazarenko,Lvov01} has also been
used to study stratified turbulence, although only considering the
coupling between the waves and not with the eddies. While this theory
can give information on the formation of the nonlinear energy cascade,
it cannot characterize interactions of the wave field with a mean flow.
Some extensions that consider nonlocal interactions and the development
of zonal flows have been proposed to address this, and are specially
relevant in the context of quasi-geostrophic turbulence \cite{James87},
and plasma turbulence in tokamaks \cite{Diamond05,Connaughton11} to
explain the development of large-scale flows.

In this article, we identify a mechanism working within a stably
stratified turbulent flow that couples the wave field with the
large-scale VSHW, thus gaining a better understanding of the role of
waves and the generation of large structures in turbulent flows. We
obtain for the first time the four-dimensional space and time resolved
energy spectra, allowing us to uniquely identify the wave components of
the total flow and to study their dynamics. We show that waves are
Doppler shifted by the VSHW, and that while a direct nonlinear energy
cascade is present, energy can also be transferred nonlocally from the
small-scale waves to the large-scale flow though CL instabilities.
Finally, we also show that the energy spread created by the Doppler
shift is not uniform, a fact that results from the turbulent nature
of the dynamically evolving large-scale flow.

\section{Stably stratified flows}

In this section we present the equations we solve numerically that
describe a stably stratified flow, introduce characteristic
time scales, and give a brief introduction to the linear theory of 
gravity waves in the presence of background shear.

\subsection{The Boussinesq equations} 

We consider the Boussinesq equations describing an incompressible 
stratified flow,
\begin{eqnarray}
    \label{bqmodel}
    \partial_t \mathbf{u} + {\bf u} \cdot {\bf \nabla} {\bf u} &=& - {\bf
        \nabla} p - N \theta \hat{z} + \nu \nabla^2 {\bf u} + {\bf F}, \\
    \label{bqmodel2}
    \partial_t \theta + {\bf u} \cdot {\bf \nabla} \theta &=& N u_z
        + \kappa \nabla^2 \theta,
\end{eqnarray}
where ${\bf u}$ is the velocity field (with 
${\bf \nabla} \cdot {\bf u} = 0$), $\theta$ is the potential 
temperature fluctuations, $p$ is the pressure normalized by 
the mean fluid density, $\nu$ is the kinematic viscosity, 
$\kappa$ is the thermal diffusivity, $N$ is the 
Brunt-V\"{a}is\"{a}l\"{a} frequency (associated with the vertical 
background stratification), and ${\bf F}$ is an external mechanical 
forcing.

In the absence of a background flow, of forcing (${\bf F}=0$), and 
of viscosity and diffusion ($\nu = \kappa=0$), Eqs.~(\ref{bqmodel}) 
and (\ref{bqmodel2}) have as exact nonlinear solutions \cite{Majda} 
internal gravity waves with dispersion relation
\begin{equation}
    \omega_0 = N \frac{k_\perp}{k} .
    \label{reldisp}
\end{equation}
Here $k_\perp = (k^2_x + k^2_y)^{1/2}$ is the wavenumber perpendicular
to gravity. Gravity acts along the $\hat{z}$ axis, with associated 
wavenumber $k_\parallel = k_z$ such that 
$k=(k^2_\perp + k^2_\parallel)^{1/2}$. 

\subsection{Characteristic timescales} 
\label{charactime}

The important characteristic timescales that come into play in this 
system are the wave period $\tau_\omega$, and the nonlinear turnover 
time $\tau_{NL}$. The wave period is simply given by 
$\tau_\omega \propto 1/\omega_0$. The nonlinear turnover time is 
the time taken by an eddie at a certain scale to transfer its energy 
through nonlinear interactions to smaller scale eddies locally in 
wavenumber space. From dimensional analysis, this timescale can 
be estimated as
\begin{eqnarray}
    \tau_{NL} \propto \frac{\ell}{u_\ell} \propto \frac{1}{k \sqrt{k E(k)}} ,
\end{eqnarray}
where $\ell \propto 1/k$ is a lengthscale in the inertial range,
$u_\ell$ is the characteristic velocity of eddies at scale $\ell$, and
$E(k)$ is the energy spectrum. If the energy spectrum in stratified
turbulence follows a power law $E(k) \propto k^{-3}$ (see, e.g.,
\cite{Billant01}), then it is easy to show that the turnover time is
independent of the scale, and becomes $\tau_{NL} \propto
L/u_\textrm{rms}$ at all scales.

\subsection{Gravity waves in a background flow} 

In the presence of a background horizontal flow ${\bf U}(z)$, the
frequency measured by an observer is shifted by the Doppler effect,
resulting in a frequency
\begin{equation}
\omega = \omega_0 + {\bf U} \cdot {\bf k}_\perp .
\label{dopplershifted}
\end{equation}
Considering for simplicity the unidirectional case with ${\bf U} = U(z)
\hat{x}$, and substituting ${\bf u} = {\bf U} + {\bf u'}$ in
Eqs.~(\ref{bqmodel}) and (\ref{bqmodel2}) with $ {\bf u'} = w(z) e^{i
    k_\perp (x-c t)} \hat{z}$ a planar internal gravity wave propagating
in $\hat{x}$ with phase velocity $c=\omega/k$, we obtain after
linearization an equation for the vertical velocity amplitude
\cite{Booker67}
\begin{equation}
   \frac{\partial^2 w}{\partial z^2} + \left[ \frac{N^2}{(U-c)^2} -
        \frac{d^2 U/d z^2}{U-c} -k_\perp^2
      \right] w = 0 .
    \label{bookereq}
\end{equation}
This equation has a singularity when $c=U(z)$, which is associated with
the existence of a CL in which the Reynolds stress tensor has a
discontinuity \cite{Booker67}. As a result, a wave traveling through the
fluid an approaching the layer with $c=U(z)$ develops faster and faster
fluctuations in its vertical velocity amplitude, becomes unstable, and
finally degenerates into turbulence, transferring a fraction of its
energy to the background flow, while the rest is dissipated. These
latter processes cannot be properly described by the linear theory. In 
practice, the instability only occurs if the Richardson number, 
$\textrm{Ri} = N^2/(\partial_z U)^2$, is greater than 
$1/4$ \cite{Booker67}.

Going beyond the linear regime, the detailed numerical study in
\cite{Winters94} considers the energy balance when a wave packet 
crosses a CL. The authors were able to quantify that more than a third
of the energy in the waves is nonlocally transferred to the mean 
background flow, while the rest of the energy in the waves is
transferred towards the small scale turbulence where it is finally
dissipated.

\section{Numerical simulations} 

Direct numerical simulations of Eqs.~(\ref{bqmodel}) and
(\ref{bqmodel2}) were performed in a cubic periodic box using the
parallel code GHOST \cite{Gomez05a,Gomez05b,Mininni11} that uses a
pseudospectral method to estimate spatial derivatives, and evolves the
system in time using a second-order Runge-Kutta method.  Two forcing
functions were used: Taylor-Green forcing with ${\bf F} = f_0(\sin x
\cos y \cos z \, \hat{x} -\cos x \sin y \cos \, \hat{z})$, and a
randomly generated isotropic three-dimensional forcing acting at $k=1$.
As we will consider frequency spectra, in both cases the forcing was
kept constant in time to prevent exciting spurious timescales in the
system, and to prevent disrupting the development of VSHW.

A Taylor-Green flow excites two-dimensional motions and has been used 
in previous studies of stably stratified turbulence to mimic atmospheric
motions \cite{Riley03}, while isotropic forcing injects energy directly
into vertical motions and thus excites stronger gravity waves. Some of
the effects we show are disrupted when VSHW cannot fully develop, so
in the next section we focus first on the isotropically forced case
with strong stratification. Then, in Sec.~\ref{comparison} we 
present results comparing both forcings and explain in more detail the 
impact of Taylor-Green forcing in the development of the CL
instability. In that section, we also analyze the effects of varying
the level of stratification.

The simulations can be characterized with dimensionless numbers. All
simulations have Reynolds number $\textrm{Re} = Lu_\textrm{rms}/\nu
\approx 9700$ (where $L$ is the forcing scale and $u_{rms}$ is the
r.m.s.~velocity), and Prandtl number $\textrm{Pr} = \nu/\kappa = 1$. For
each forcing function, we performed three simulations with varying
stratification $N$, resulting in Froude numbers $\textrm{Fr} =
u_\textrm{rms}/(NL) \approx 0.04$, $0.02$, and $0.01$ (with decreasing
$\textrm{Fr}$ indicating stronger stratification), and buoyancy Reynolds
number $\textrm{Re}_\textrm{b} = \textrm{Re}\textrm{Fr}^2 \approx 15$,
$3.9$ and $0.97$. Each simulation was started from the fluid at rest, so
neither the large scale flow nor the small scale waves were previously
imposed, and the simulations were continued for $12$ large-scale
turnover times to let the system reach a turbulent steady state. Then,
each simulation was continued for other $12$ turnover times saving all
fields with high temporal cadence, resulting in 1200 outputs of each
field that sample the Brunt-V\"{a}is\"{a}l\"{a} frequency 40 times per
period when $\textrm{Fr} \approx 0.01$. This, previously unmatched
temporal resolution, allows us to resolve four-dimensional spectra
simultaneously in space and in time, although at a computational cost
that allows us to reach only moderate spatial resolutions with $512^3$
spatial grid points. Nonetheless, as we show, the spatial resolution we
use is appropriate, and recent studies also indicate that properties of
fully developed turbulence can be identified even at moderate
resolutions \cite{Schumacher14}.

\begin{figure}
    \centering
    \includegraphics[width=8.5cm]{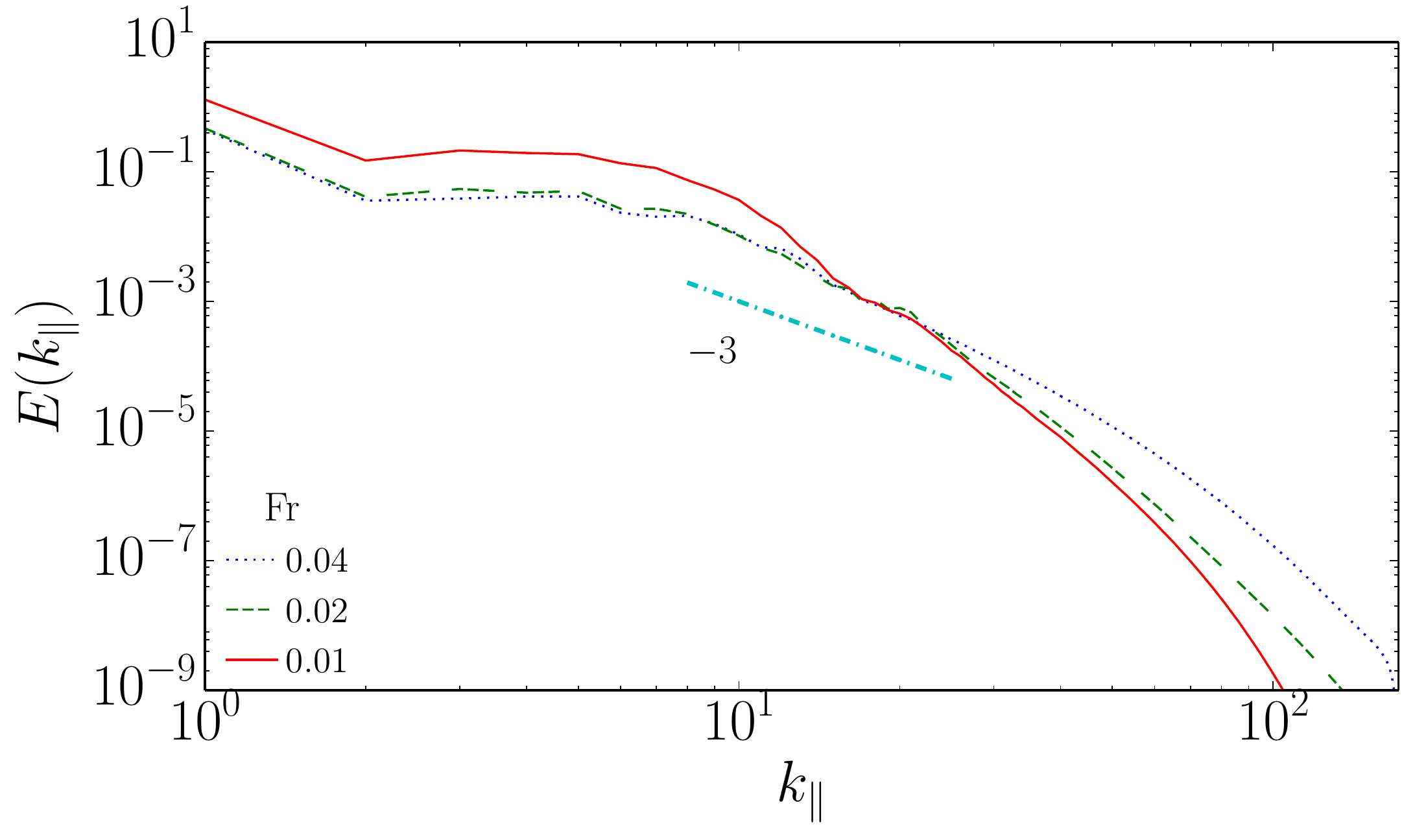} 
    \caption{{(\it Color online)} Reduced parallel energy spectrum
        $E(k_\parallel)$ for the three simulations with isotropic
        forcing. A $k_\parallel^{-3}$ slope is shown only as a
        reference.}
    \label{reduced}
\end{figure}

\begin{figure}
    \centering
    \includegraphics[width=7cm]{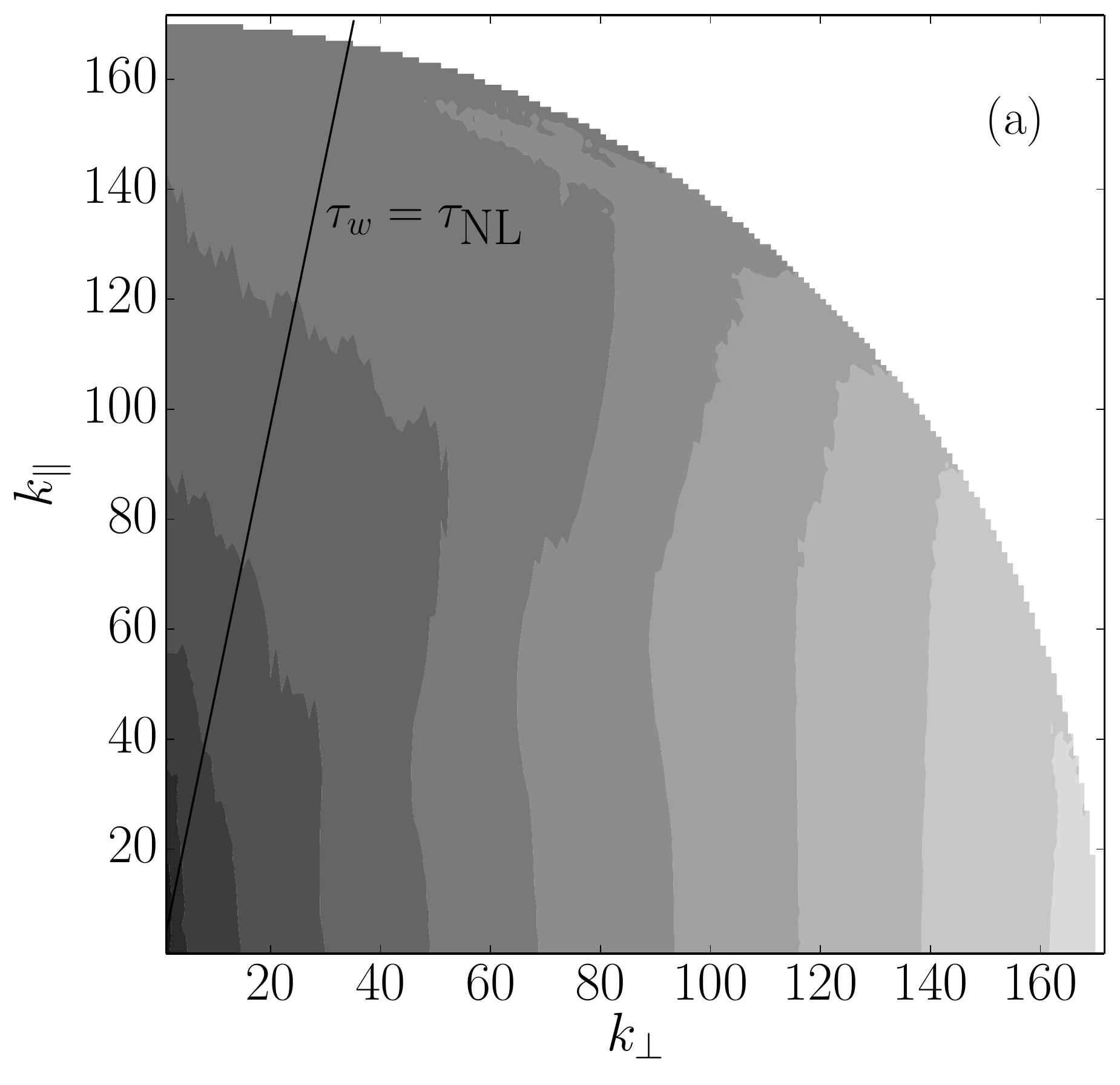} \\
    \includegraphics[width=7cm]{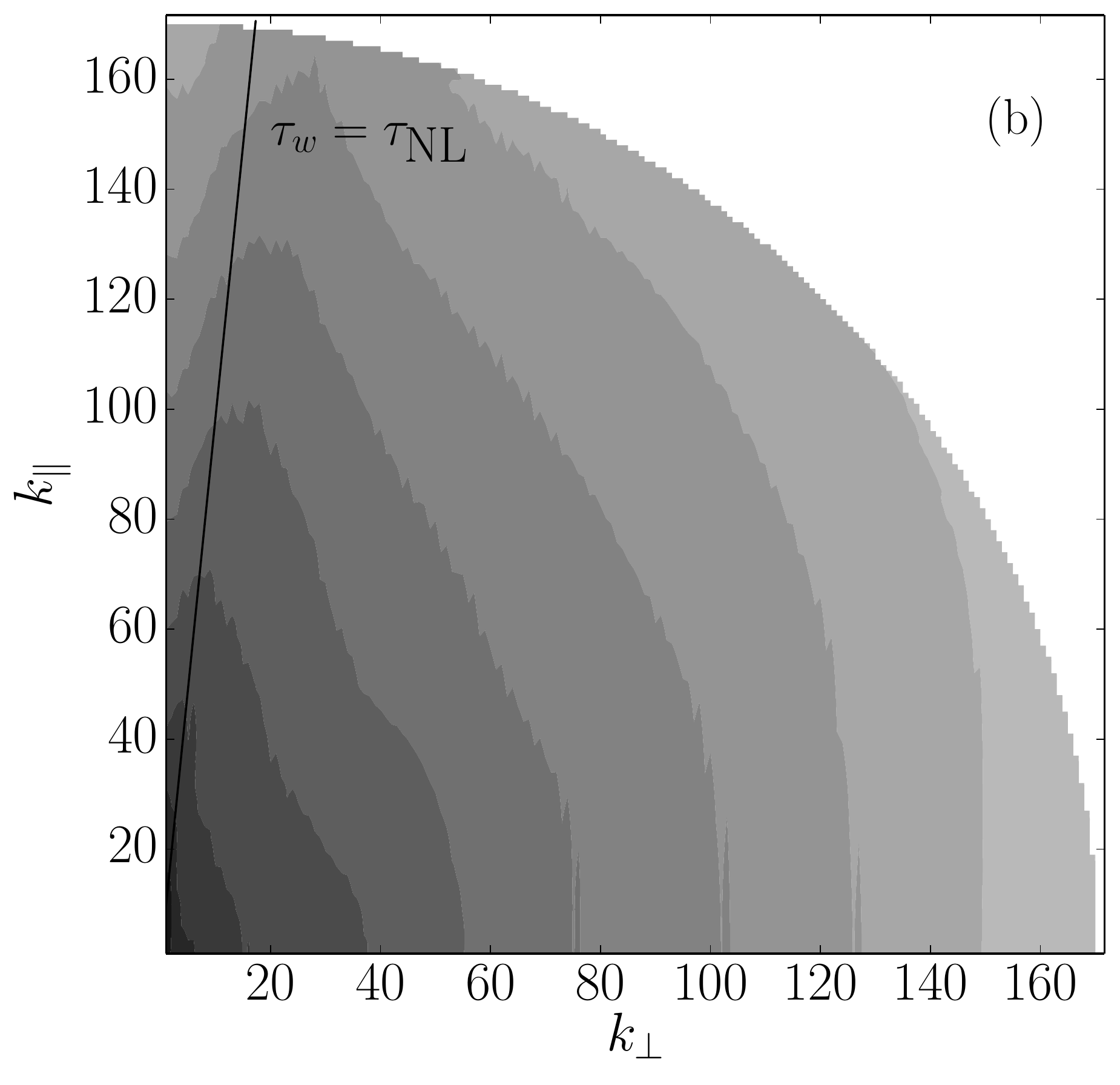}
    \caption{{(\it Color online)} Space resolved axisymmetric energy spectrum 
        $e(k_\perp,k_\parallel)$ for two runs with (a) $\textrm{Fr}
        \approx 0.02$, and (b) $\textrm{Fr}\approx 0.01$. Note the 
        spectral anisotropy resulting from stratification. The solid
        line indicates the modes with wave period equal to the eddy 
        turnover time, $\tau_\omega = \tau_\textrm{NL}$. In (b) a
        ridge is formed close to this curve, indicating energy is
        transferred towards modes with lower $k_\perp$ but the transfer 
        is halted when $\tau_\omega \approx \tau_{NL}$. Modes with 
        $\tau_\omega < \tau_\textrm{NL}$ (i.e., modes below the solid 
        line) are often called wave modes, as these modes have the 
        wave period as the fastest timescale. However, a proper
        characterization of waves requires space and time resolved 
        spectra.}
    \label{kspec}
\end{figure}

\section{Results} 

\subsection{Spatial spectral analysis} 

Traditional characterization of turbulent flows is done using one-dimensional
spectra. In Fig.~\ref{reduced} we present the parallel spectrum $E(k_\parallel)$
for the three simulations with isotropic forcing. Only as a reference, we also
show in Fig.~\ref{reduced} a power law $\propto k_\parallel^{-3}$. The reader
can find detailed spatial characterizations of stratified turbulence, which go
beyond the aim of this work and at higher spatial resolution, in
recent studies (see, e.g., \cite{Waite06,Brethouwer07}).

As a better way to characterize energy distribution among scales in
the presence of anisotropy, Fig.~\ref{kspec} shows the spatial 
axisymmetric energy spectrum $e(k_\perp,k_\parallel)$, normalized by 
$\sin \theta$ with $\theta = \arctan (k_\perp/k_\parallel)$ to obtain 
circular isocountours in the case of an isotropic flow. In 
Fig.~\ref{kspec}, two spectra are presented, for the simulations 
with isotropic forcing and $\textrm{Fr} \approx 0.02$ and 
$\approx 0.01$.  As stratification is increased, energy distribution 
becomes more anisotropic, with energy being preferentially transferred 
towards modes with smaller $k_\perp$ (and, as a result, larger wave period)
\cite{Billant01,Waite06,Brethouwer07,Smith02}. However, for $\textrm{Fr}
\approx 0.01$ energy accumulates near the modes with wave period
($\tau_\omega \propto 1/\omega_0$) equal to the nonlinear turnover time
($\tau_\textrm{NL} \propto L/u_\textrm{rms}$), forming a ridge. As the
energy transfer mechanism is often given by the shortest timescale
\cite{Clark14}, modes below the curve $\tau_\omega = \tau_\textrm{NL}$
(those with wave period shorter than the turnover time) are associated
with waves \cite{Smith02,Waite06,Brethouwer07}. Modes above the curve
$\tau_\omega = \tau_\textrm{NL}$ (and in particular, modes with
$k_\perp=0$) are often called vortical modes, as for these modes
$\omega_0 \approx 0$.  The fraction of the energy in these modes
decreases with decreasing $\textrm{Fr}$, in good agreement with the
observed accumulation near $\tau_\omega = \tau_\textrm{NL}$ for large
$\textrm{Fr}$.  The slow down of the transfer as the energy reaches the
ridge is compatible with critical balance arguments \cite{Nazarenko},
and is also of great importance in weak wave turbulence theories
\cite{Lvov01} that require the energy of the system to remain in weakly
interacting waves.  Note however that this distinction between waves and
vortical modes in Fig.~\ref{kspec}, based on the characteristic time 
of each mode, is only approximate. A strict discrimination between 
waves and eddies requires the four-dimensional spectrum.

\subsection{Spatio-temporal analysis} 

\begin{figure}[h!]
    \centering
    \includegraphics[width=8.5cm]{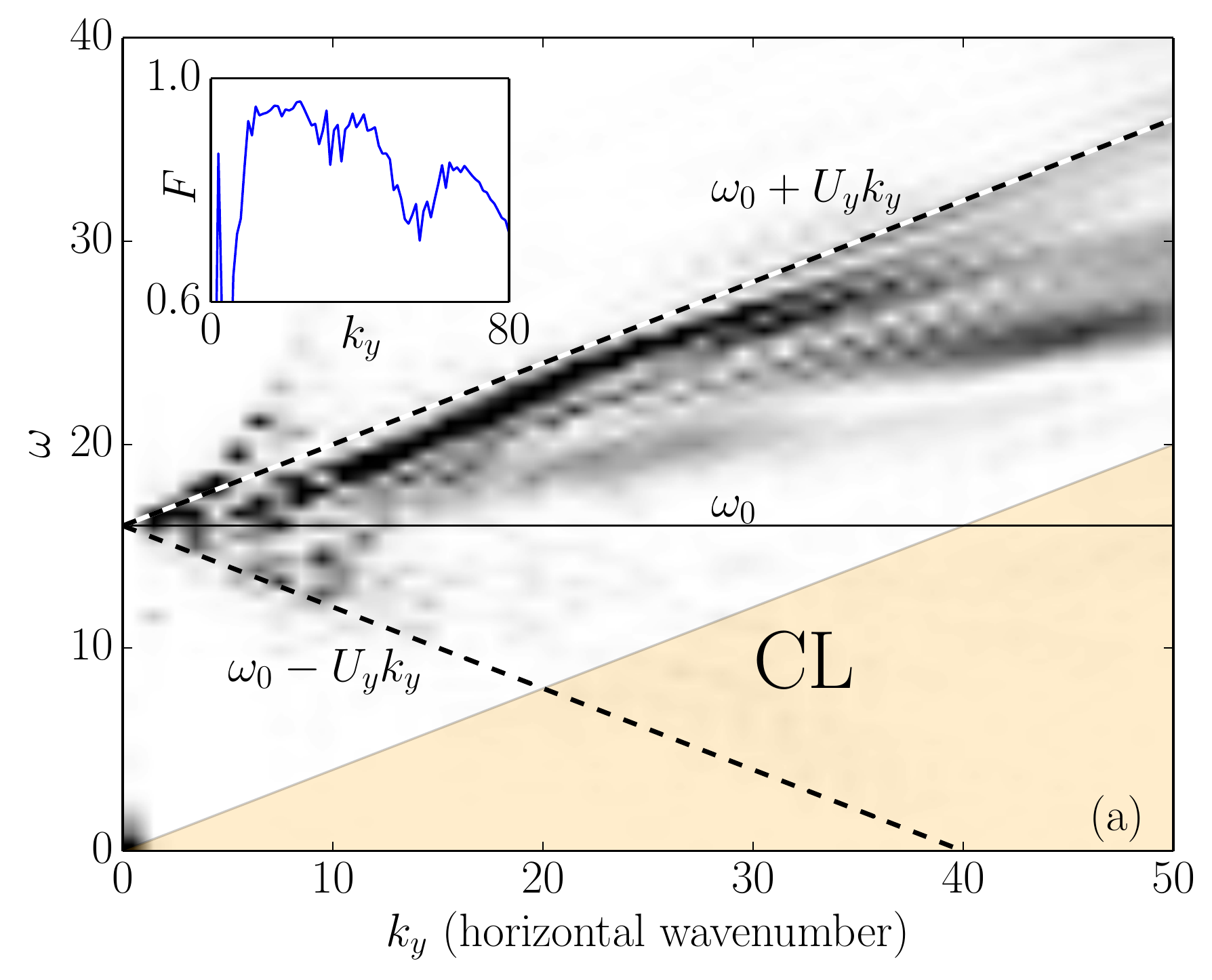} \\
    \includegraphics[width=8.5cm]{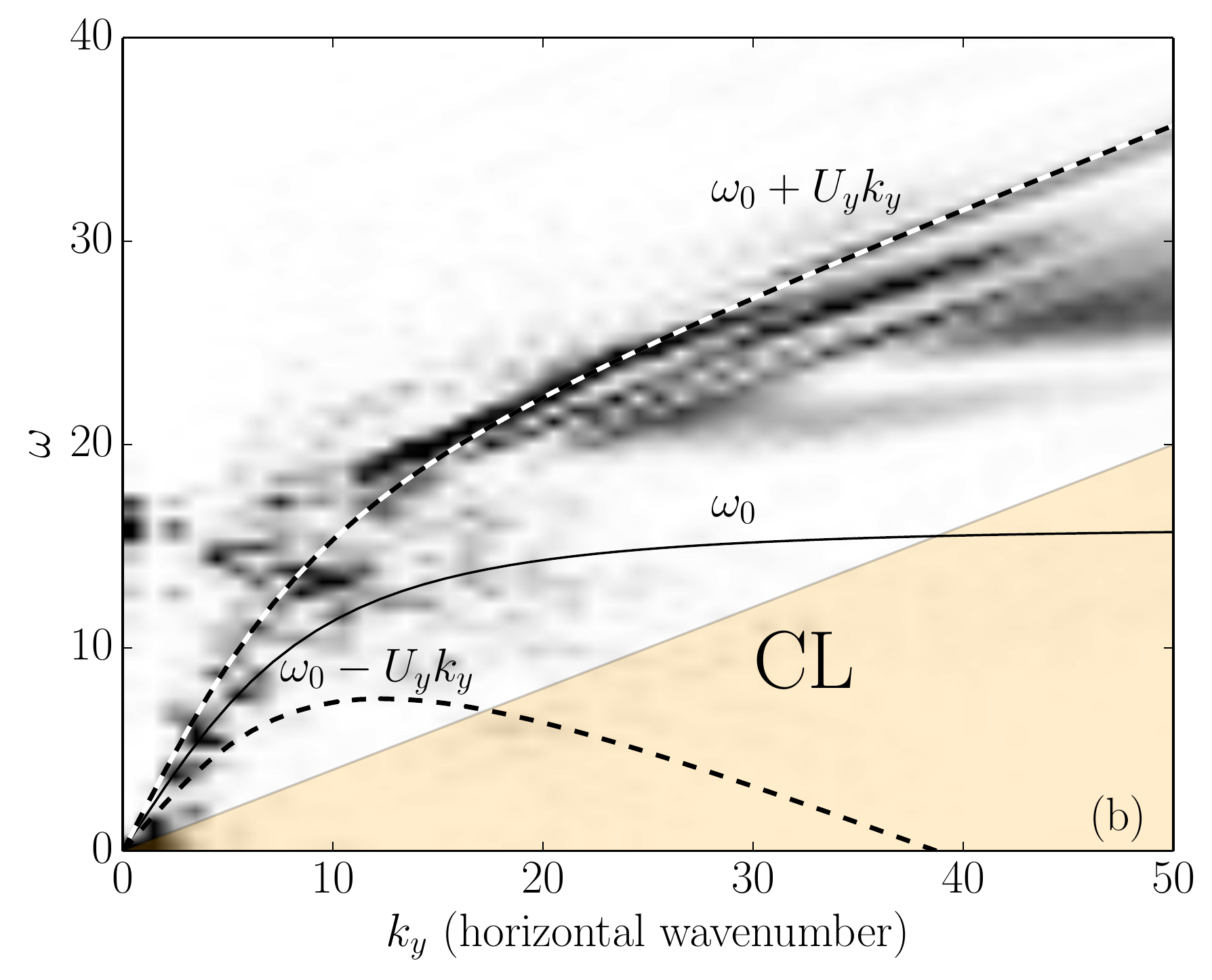}
    \caption{{(\it Color online)} Space and time resolved spectrum of
        the potential energy $E_\theta(k_x=0,k_y,k_z,\omega)$
        (normalized by $E({\bf k})$, with 0 corresponding to white and 
        $1$ corresponding to black), for $\textrm{Fr} = 0.01$, and for 
        two different values of $k_z$: (a) $k_z = 0$, and (b) $k_z = 10$.
        The fundamental dispersion relation $\omega_0({\bf k})$ from 
        Eq.~\eqref{reldisp} is given by the thin solid curve, along 
        with two Doppler shifted branches with $U_y= \pm 0.4$ (dashed
        curves). Energy is mostly concentrated in the fan defined by
        the two shifted branches, although not uniformly distributed. 
        Waves that travel with the flow (upper half fan) concentrate most 
        of the power. The area shaded with transparent orange
        corresponds to $\omega < U_y k_y$ with $U_y=0.4$; note there
        is almost no power in this region. The defect of energy in all 
        modes in this area indicates these waves are absorbed by 
        critical layers (CL), with their energy being transferred to the flow. 
        {\it Inset:} Fraction of the energy $F$ that is contained within 
        the two Doppler shifted branches as a function of the wavenumber. 
        In the inertial range, $\approx 80\%$ of the energy corresponds
        to Doppler shifted waves.}
    \label{ekw}
\end{figure}

\begin{figure}
    \centering
    \includegraphics[width=8cm]{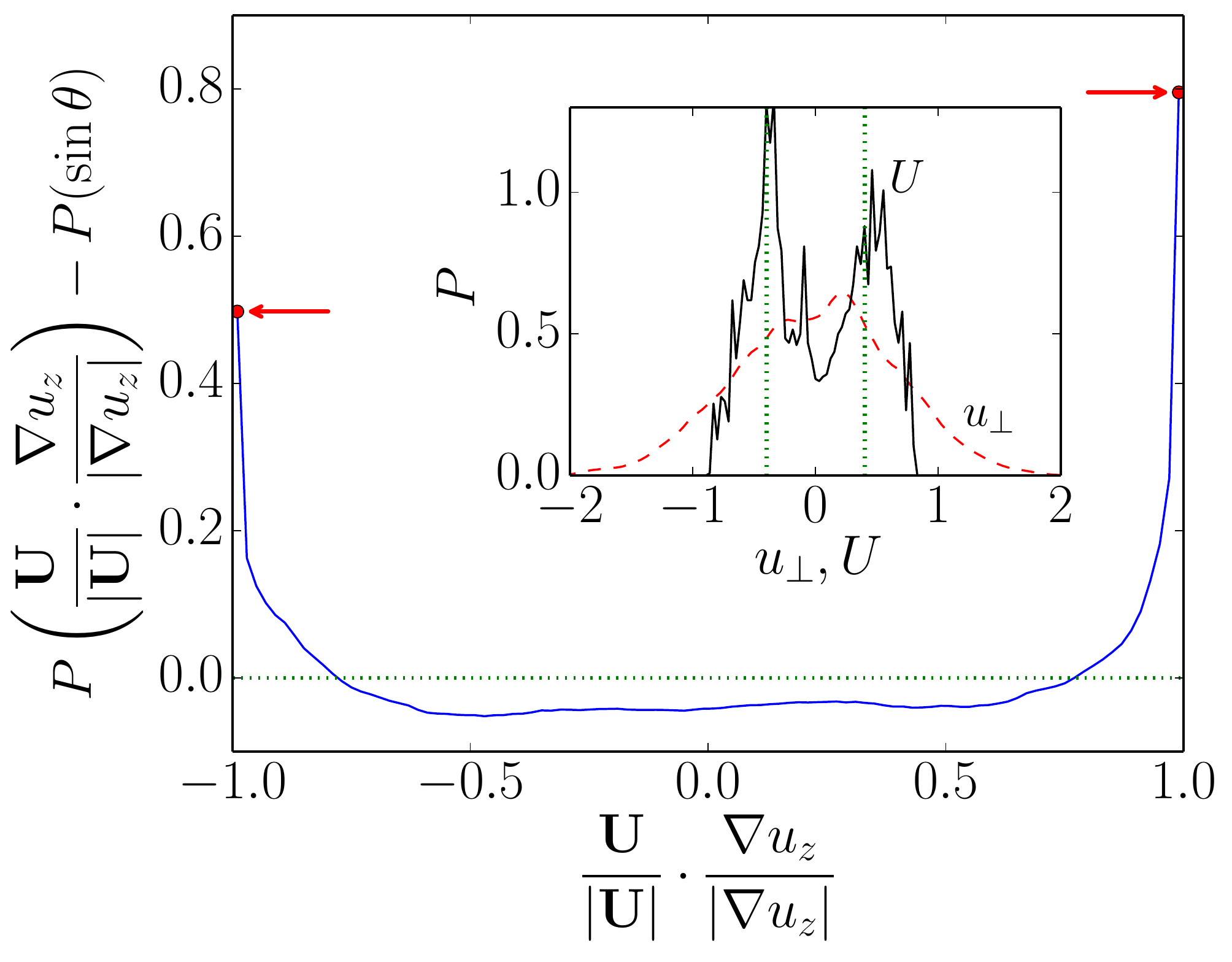} 
    \caption{{(\it Color online)} Probability density function for
        horizontal gradients of the vertical velocity being aligned or
        anti-aligned with the mean horizontal flow, minus the
        probability of having uniformly distributed alignment between
        the fields ($\theta$ is uniformly distributed in $[0,2\pi)$).
        The PDF indicates (independently of Fig.~\ref{ekw}) that there
        is an excess of waves traveling with the mean horizontal flow
        (compared with against the flow), and also a defect of waves
        propagating perpendicular to the flow.  {\it Inset:} Probability
        distribution of the Cartesian components of the mean horizontal
        velocity ${\bf U}$, and of the point-wise horizontal velocity
        ${\bf u}_\perp$. The dotted vertical lines indicate the values
        $\pm 0.4$, used to draw the Doppler shifted branches in
        Fig.~\ref{ekw}.}
    \label{check}
\end{figure}

Precise identification of the waves, and of their role in the dynamics,
requires both space and time information. Figure \ref{ekw} shows
different cuts of the frequency and wavenumber spectrum $E_\theta
(k_x,k_y,k_z,\omega)$ for $k_x=0$, and for either $k_z=0$ or $k_z=10$
in the simulation with the strongest stratification and isotropic forcing.
As internal gravity waves couple vertical motions with temperature
fluctuations, we consider the spectrum of potential energy $E_\theta$ to
isolate the waves more easily. In Fig.~\ref{ekw} there is no significant
accumulation of energy in the modes that satisfy the dispersion relation
given by Eq.~(\ref{reldisp}), i.e., in modes that could be directly
associated with internal gravity waves.  Instead, most of the energy
($\approx 80\%$ for $k_y \in [10,80]$) falls inside a wider region
defined by the Doppler shifted Eq.~\eqref{dopplershifted} with $-0.4 \le
U_y \le 0.4$. This suggests the waves are propagating through layers
with non-zero mean horizontal velocity (i.e., VSHW). Indeed, a
probability density function (PDF) of the Cartesian components of the
horizontally averaged horizontal velocity ${\bf U}$ obtained directly
from the same simulation shows two peaks at $\approx \pm 0.4$ (see the
inset in Fig.~\ref{check}). Also, the PDF of the point-wise horizontal
velocity in the entire box is approximately Gaussian with variance close
to $0.4$.

It is important to remind that we force the fluid from rest at
the large scales; this generates the winds, and by nonlinear
interactions energy cascades towards the smaller scales. Then, the cycle
completes itself when the small scale waves interact with the large
scale winds; we are not imposing flows at different scales, or
imposing a background flow, as sometimes done to study interaction 
of waves with shear flows. Instead, we just let turbulent interactions
develop by themselves. In similar simulations of turbulent stratified 
flows, waves have been directly observed before by measuring the 
frequency spectrum $E(\omega)$, or by measuring $E(\omega)$ for a 
few Fourier modes \cite{Lindborg07}. This study identified peaks at 
the frequency of the internal gravity waves, but from the frequency
spectrum a relation between frequency $\omega$ and wavenumber 
${\bf k}$ cannot be obtained without assuming the system is
dominated by the waves (and then the relation of these quantities is
given by the dispersion relation), or dominated by vortical motions 
(and then the relation is given by sweeping, see, e.g., \cite{Clark14}).

Figure \ref{ekw} gives direct evidence that most of the energy is in 
waves. And the waves satisfy Eq.~\eqref{dopplershifted} where ${\bf U}$ 
is the horizontal wind. However, there is more power in modes close 
to the curve $\omega_0+U_y k_y$ (with $U_y = 0.4$) than in modes 
close to $\omega_0-U_y k_y$. This energy distribution cannot be 
explained by a preferential direction in ${\bf U}$ (see the PDFs in the inset in
Fig.~\ref{check}), by viscous effects (which introduce damping but no
frequency shift), or by nonlinear corrections to the dispersion
relation, as internal gravity waves are exact nonlinear solutions of the
Boussinesq equations \cite{Majda}. The preferential concentration is
instead compatible with a CL instability. In Fig.~\ref{ekw} we show as a
reference the area (shaded in transparent gray or orange) where 
$\omega < U_y k_y$ with $U_y = 0.4$. As in this simulation 
$-0.4 \lesssim U_y \lesssim 0.4$ (once again, see the PDFs in the inset in
Fig.~\ref{check}), for all modes in that area there is some layer with
$U_y$ such that the layer can be a CL. When the waves approach these
layers, $c \approx U$, and from Eq.~\eqref{bookereq}  their vertical
velocity starts fluctuating more rapidly, undergoing an instability and
transferring a fraction of their energy to the flow
\cite{Hines91,Winters94}  (i.e., the waves are absorbed).  Indeed, 
there is almost no energy in the wave modes in this area. 
This mechanism can only act if $\textrm{Ri} >1/4$.
We verified the point-wise value of $\textrm{Ri}$ in the simulations is
larger than $1/4$. For the run with $\textrm{Fr}\approx0.04$,
the minimum of $\textrm{Ri}$ is $\approx14$. Although this value is 
rather large for atmospheric flows, it has been reported in other
simulations of stratified turbulence, and indicate that CL instability 
can be the reason for the nonlocal formation of large scale structures
in stratified turbulence observed in \cite{Smith02,Marino13,Marino14}.

\begin{figure}
    \centering
    \includegraphics[width=8.5cm]{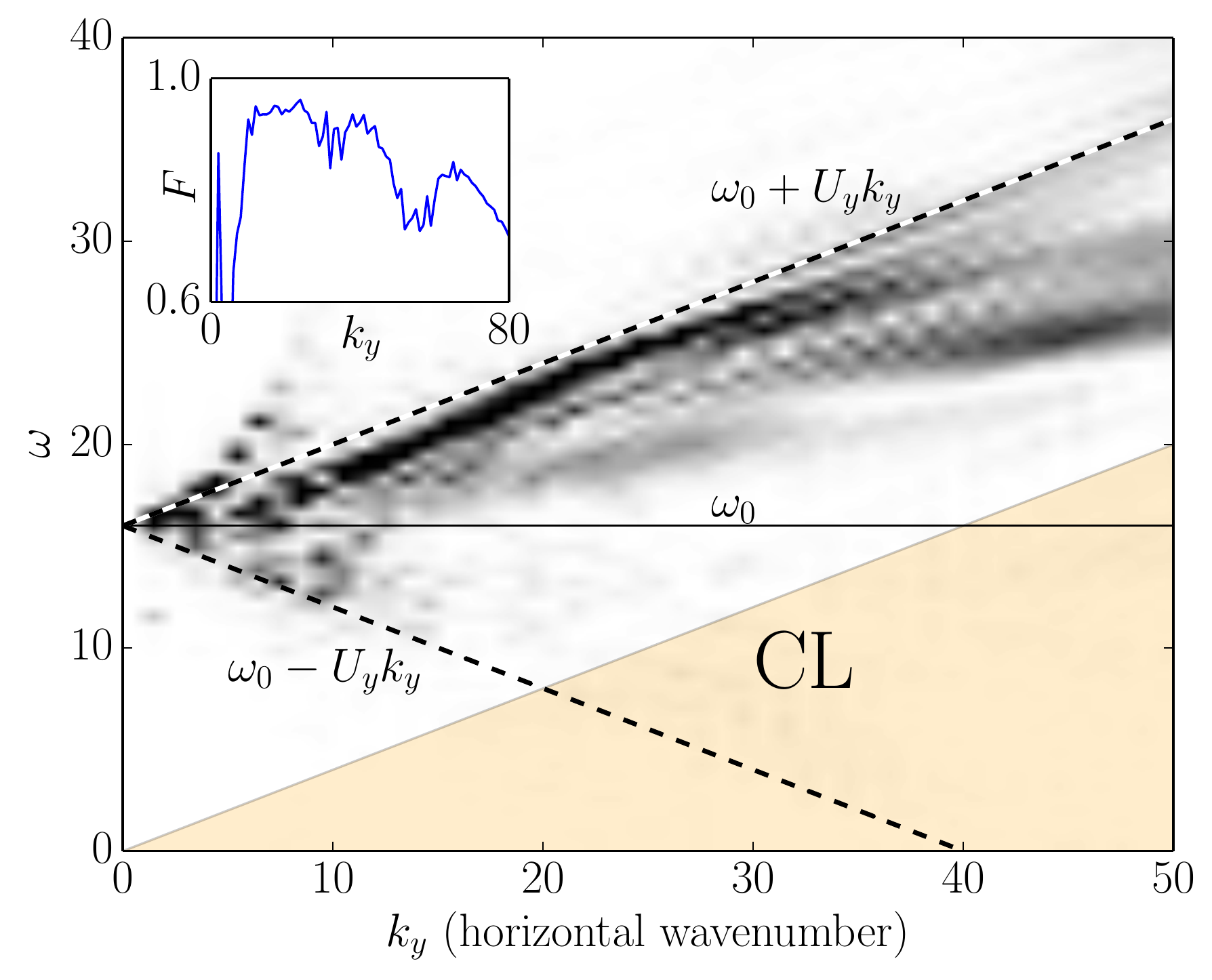} 
    \caption{{(\it Color online)} Space and time resolved spectrum of
        vertical kinetic energy $E_z(k_x=0,k_y,k_z=10,\omega)$, 
        normalized by $E_z({\bf k}$), and for $\textrm{Fr} =
        0.01$. There is no discernible difference with the space 
        and time spectrum of the potential temperature (see 
        Fig.~\ref{ekw}). The solid line corresponds to the linear 
        dispersion relation of gravity waves, while the dashed lines 
        indicate the two Doppler shifted dispersion relations with 
        $U_y= \pm 0.4$. The region shaded with transparent orange 
        indicates the modes that can have a CL instability, note the
        lack of energy in that region. {\it Inset:} Fraction of the
        energy $F$ that is contained within the two Doppler shifted 
        branches as a function of the wavenumber.}
    \label{vz}
\end{figure}

Internal gravity waves in a stratified fluid couple the temperature with
the vertical component of the velocity. We now show that the
four-dimensional power spectrum of the vertical velocity displays the
same features as the spectrum of the potential energy shown above. 
Figure \ref{vz} shows a cut of the power spectrum of the
vertical velocity $E_z(k_x=0,k_y,k_z=0,\omega)$ in the simulation
with stronger stratification and isotropic forcing. The same features
found in the four-dimensional spectrum of the potential energy can be
found in this figure, including the Doppler shift and the defect of
energy in the modes that have frequency compatible with CL
instability. In fact, the spectra are practically indistinguishable. 
As in the case of the temperature, a large fraction of the energy in
vertical motions is also associated with wave motions, ($\approx 80\%$
of the energy between $10<k_y<80$ is inside the fan corresponding to
Doppler shifted waves, see the inset in Fig.~\ref{vz}). When the
spectrum of  horizontal velocity is considered instead, none of these
signatures can be found (not shown).

As a final and independent verification of the excess of waves traveling with the
flow observed in Fig.~\ref{ekw}, we resort to a  statistical analysis.
As most of the energy in the simulation with $\textrm{Fr}\approx0.01$ is
in the waves, we can assume the vertical velocity is a superposition of
traveling waves $\propto e^{i({\bf k} \cdot {\bf x} - \omega t)}$. We
can then compute ${\bf U} \cdot {\bf \nabla} u_z/(|{\bf U}||{\bf \nabla}
u_z|)$, which gives an estimation of how the wave vector (i.e., the
propagation direction for a pure planar wave) is aligned with the
horizontal flow. Figure \ref{check} shows the PDF of this alignment
(averaged over every layer in the fluid) minus the PDF of the sine of a
uniformly distributed angle $\theta \in [0,2\pi)$ (i.e., of randomly
aligned fields with no privileged direction), for the simulation
with stronger stratification and isotropic forcing. The result indicates a
preference towards an alignment of horizontal gradients of the vertical
velocity with the mean flow, as there is an excess for $+1$ when
compared with $-1$, and a deficit (compared with uniformly distributed
angles) for the case in which the two fields are perpendicular.
Interestingly, a preference of stratified flows towards developing one
sign of velocity gradients (resulting in non-Gaussian PDFs) was reported
before in \cite{Rorai13}.

An interesting fact, that sheds more light on how these mechanisms work,
is that these effects are not observed in purely rotating flows
\cite{Clark14,Yarom14}. While these flows can also generate strong
horizontal velocity fields, waves tend to travel in the vertical
direction (as compared to the stratified case, where they tend to travel
horizontally), and no Doppler shift develops as the waves are
perpendicular to the large-scale flow.

\subsection{Dependence on level of stratification and forcing}
\label{comparison}

The results discussed so far correspond to the simulation with 
randomly generated isotropic forcing and $\textrm{Fr} \approx 0.01$. 
A similar distribution of energy in frequency and wavenumber is observed
in the simulations with same forcing but with higher $\textrm{Fr}$
(i.e., weaker stratification), although the excess of waves traveling
with the flow (as well as the defect in the fan defined by the CL) is 
stronger in the flow with smallest $\textrm{Fr}$. Figures showing 
the dependence with Froude number are shown below. For Taylor-Green 
forcing, which prevents the formation of a mean wind in each
horizontal layer, the CL instability is not observed, although
Doppler spreading of the waves still takes place. These results are
also presented in this section.

\begin{figure}
    \centering
    \includegraphics[width=8.5cm]{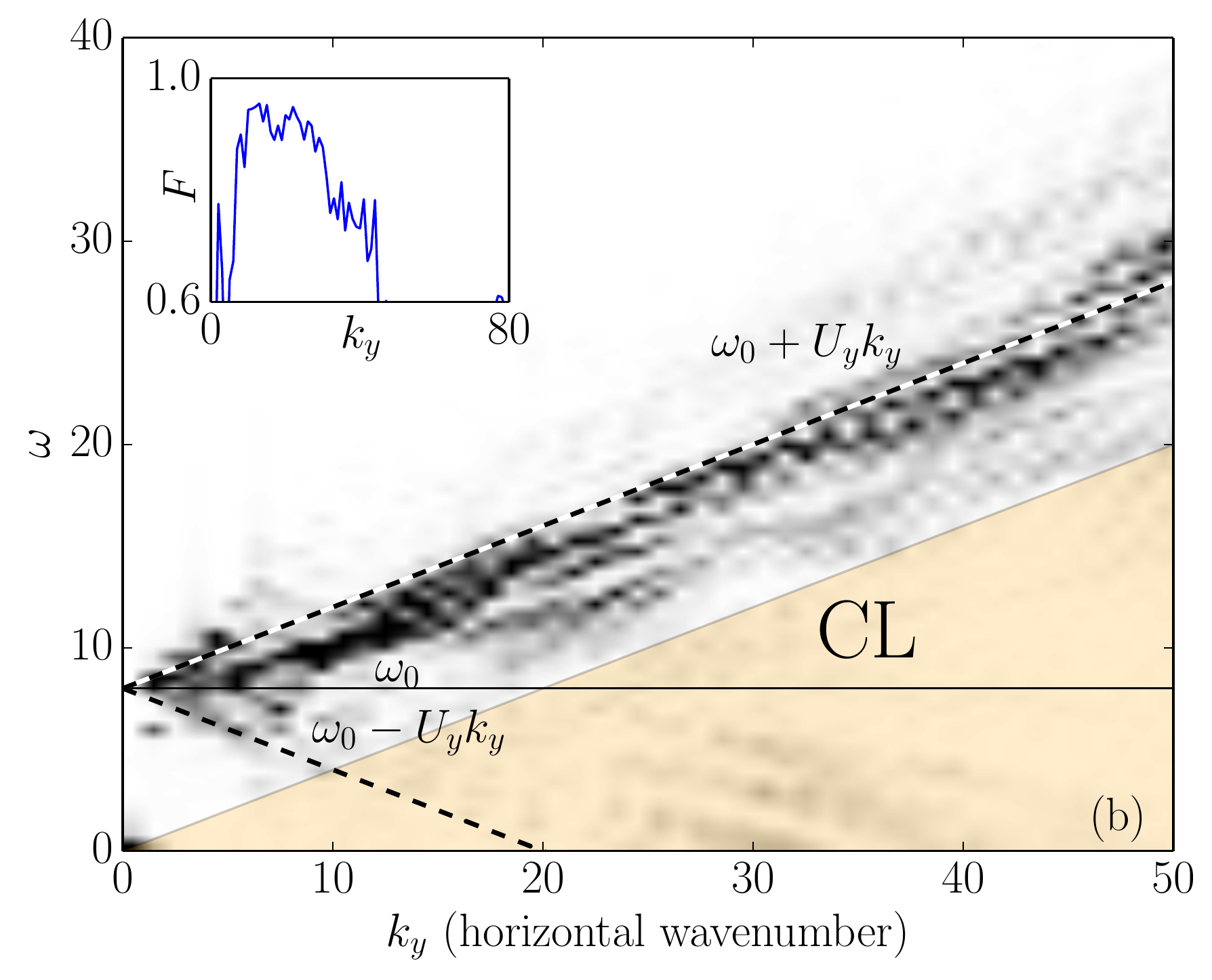} \\
    \includegraphics[width=8.5cm]{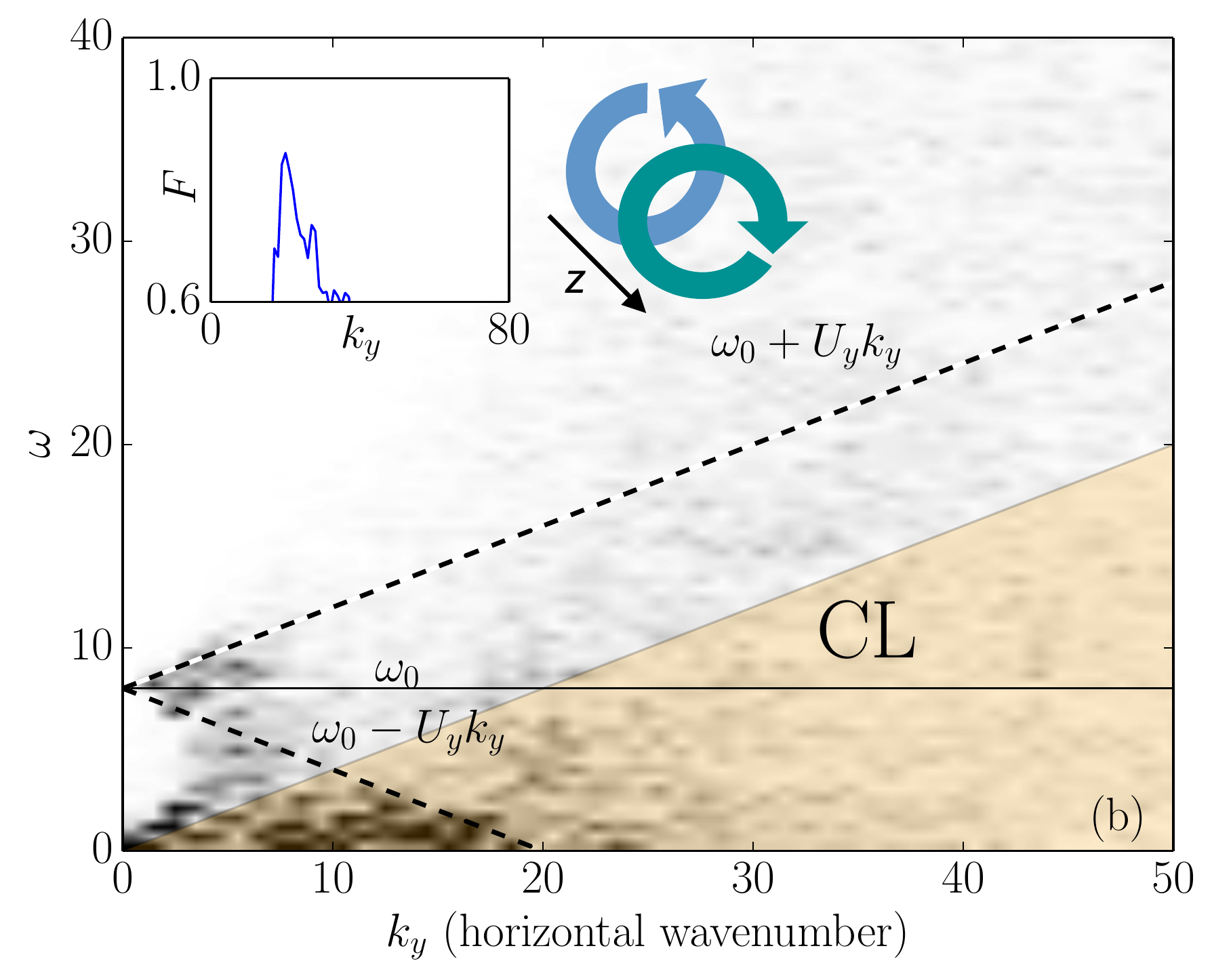}
    \caption{{(\it Color online)} Space and time resolved spectrum of
        the potential energy $E_\theta(k_x=0,k_y,k_z=10,\omega)$
        (normalized by $E_\theta({\bf k}$) for $\textrm{Fr} = 0.02$, and
        for two different forcing functions: (a) isotropic forcing, and
        (b) Taylor-Green forcing. Labels are as in Fig.~\ref{vz}. The
        insets show in each case the fraction of the energy $F$ that is
        contained within the two Doppler shifted branches.  In (b), the
        inset on the right also shows a schematic representation of the
        Taylor-Green forcing. In (a), note the decrease in the level of
        stratification of the system still preserves the main features
        of the spectrum. In (b), they Taylor-Green forcing injects most
        of the energy in vortical modes ($\omega \approx 0$), the energy
        in the waves is therefore smaller, and the flow geometry
        prevents the CL instability from developing.}
    \label{tgn}
\end{figure}

As a comparison with the case shown in Fig.~\ref{ekw},
Fig.~\ref{tgn}(a) shows the space and time resolved spectrum 
$E_\theta (k_x=0,k_y,k_z=0,\omega)$ for a simulation still 
forced with randomly generated forcing, but with 
$\textrm{Fr} \approx 0.02$.  This spectrum can be directly compared with
Fig.~\ref{ekw}(a). Decreasing the stratification results in a spectrum
that bears great resemblance with the spectrum for $\textrm{Fr} \approx
0.01$, with non-uniform Doppler spreading of the waves and the excess of
waves with $\omega \approx \omega_0 + U_y k_y$. The defect of energy in
the modes compatible with CL instability is also visible in this
simulation, although the damping of these modes is slightly weaker than
in the case with stronger stratification. This is compatible with the
smaller fraction of energy in the waves as stratification is decreased;
see the inset of Fig.~\ref{tgn}(a). Also, this is compatible with the
fact that as stratification is decreased, the power in the large-scale
motions (small wavenumbers) also decreases (see Fig.~\ref{reduced}).
These trends are further confirmed by the simulation with weaker 
stratification and $\textrm{Fr} \approx 0.04$.

When Taylor-Green forcing is used instead, the broadening of the
dispersion relation by Doppler shift can still be observed; see
Fig.~\ref{tgn}(b) for $\textrm{Fr} \approx 0.02$, specially for
wavenumbers $0<k_y \lesssim 10$. However, this effect is dimmed by a
large concentration of energy in modes with $\omega \approx 0$. This is
to be expected as the Taylor-Green forcing, given by ${\bf F} = f_0(\sin
x \cos y \cos z \, \hat{x} -\cos x \sin y \cos\, \hat{z})$, consists of
two counter rotating vortices in the horizontal velocity, and excites no
vertical motions directly (see a schematic diagram in the inset of
Fig.~\ref{tgn}(b)). As a result, the fraction of energy in wave modes is
much smaller than in the case with isotropic forcing (see also a direct
measurement of this in the inset of Fig.~\ref{tgn}(b)). Moreover, the
imposition of the Taylor-Green vortices in each layer by the external
forcing seems to disrupt the development of a non-zero mean wind in each
horizontal layer, weakening also the development of the CL instability.
Indeed, the damping of the energy in modes consistent with CL
instability in the four-dimensional spectrum of potential energy in
Fig.~\ref{tgn}(b) is almost inexistent.

\begin{figure*}
    \centering
    \includegraphics[width=8cm]{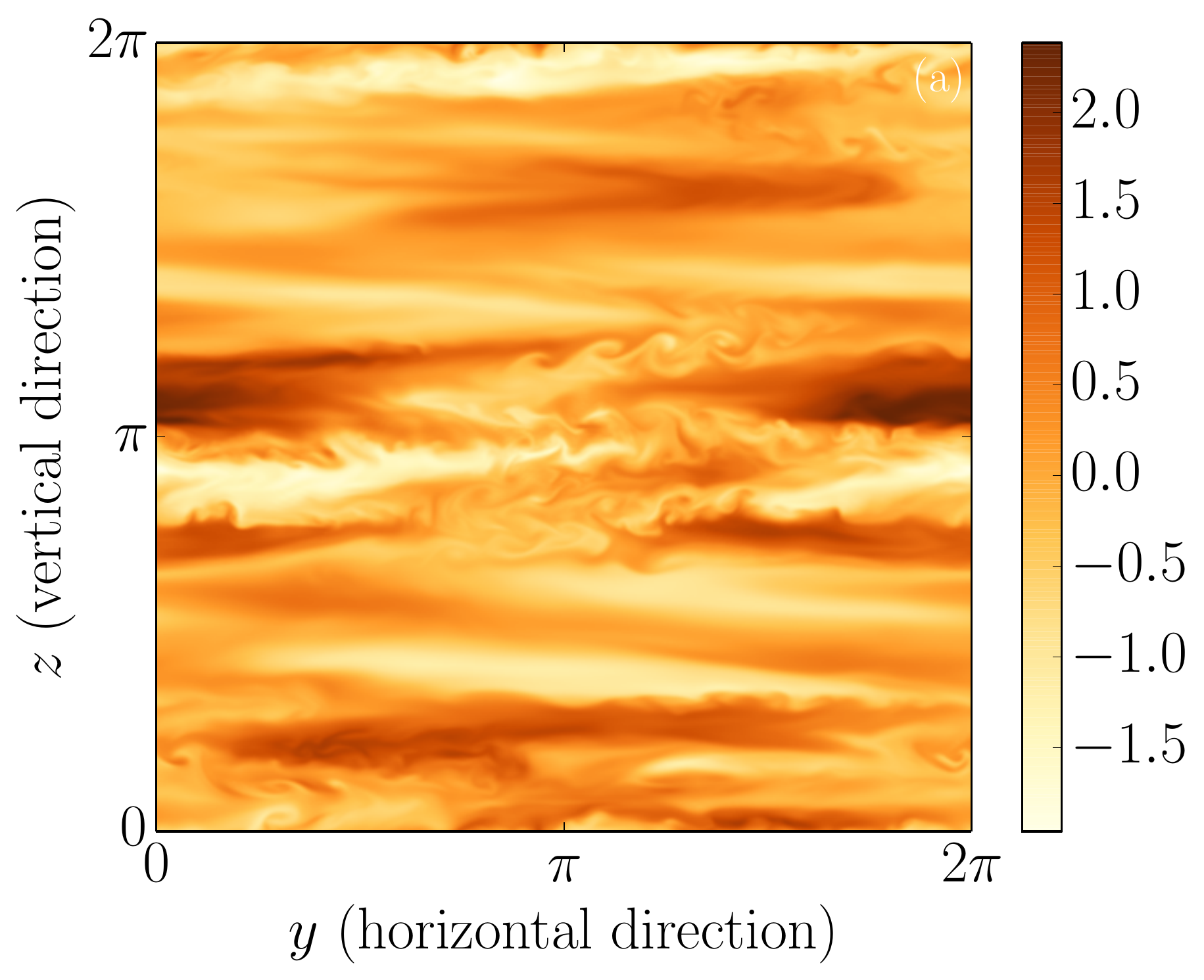}
    \includegraphics[width=8cm]{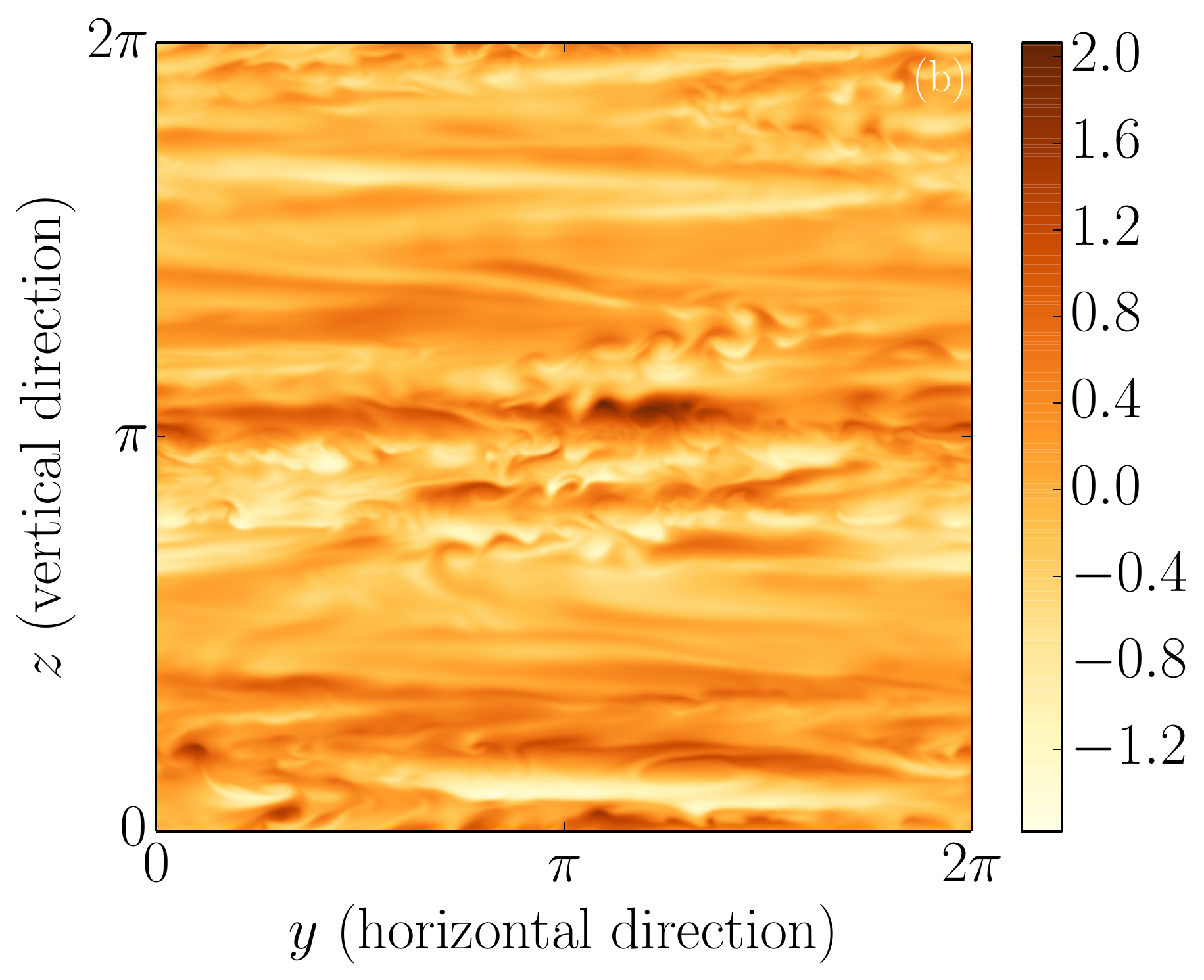}\\
    \includegraphics[width=8cm]{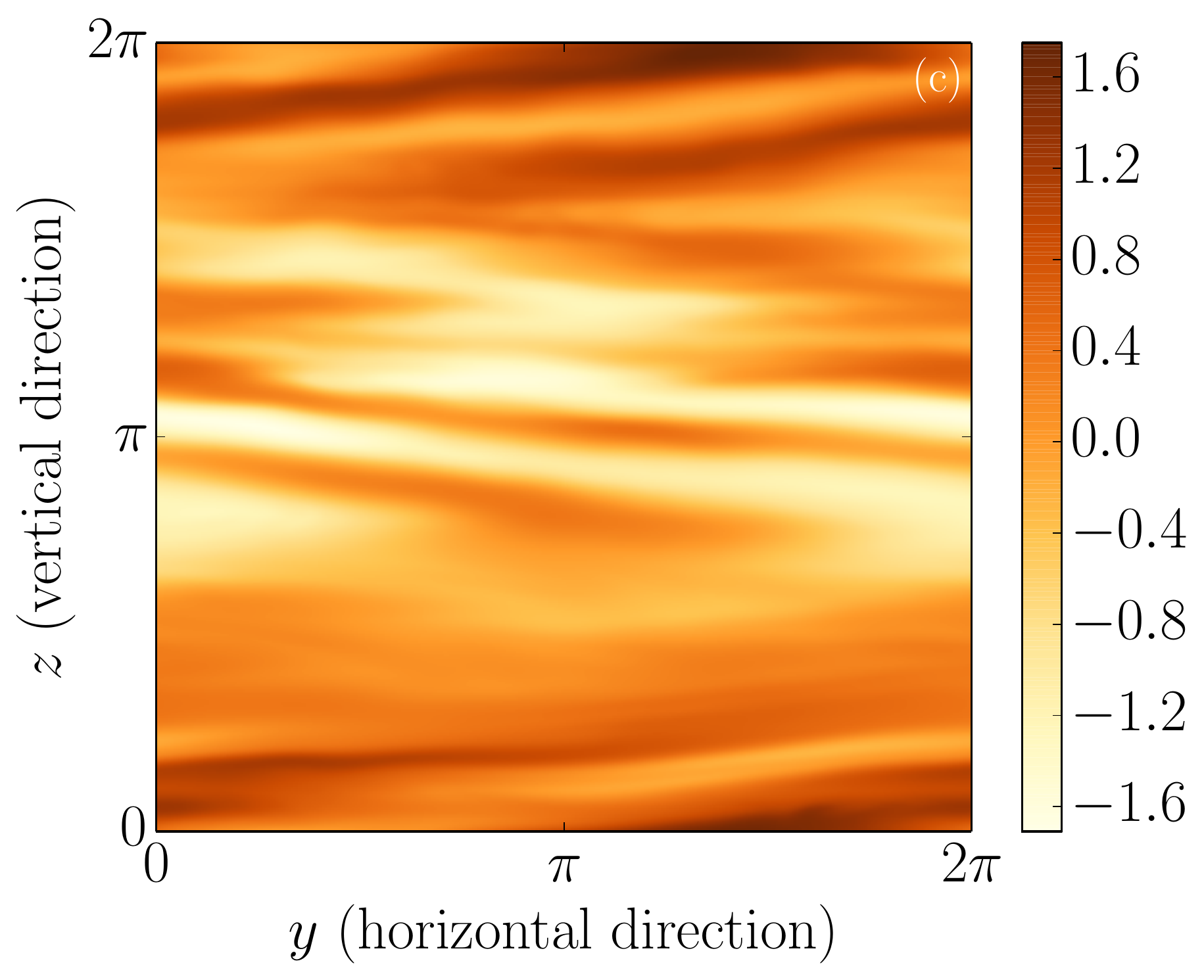}
    \includegraphics[width=8cm]{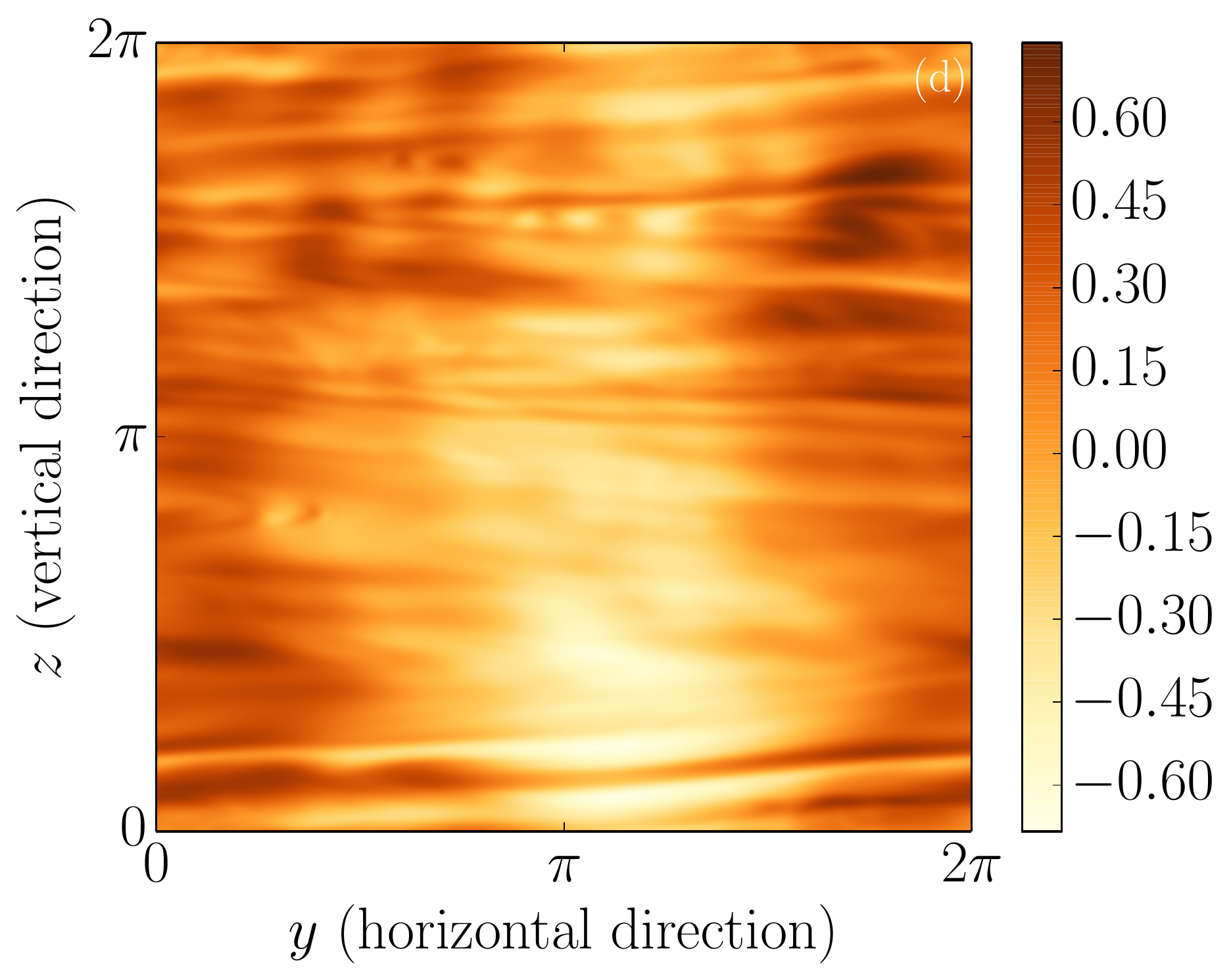}
    \caption{Vertical cuts (at constant $x$) of the potential
        temperature $\theta$, and of the $y$-component of the velocity
        field $u_y$, for simulations with the same $\textrm{Fr}
        = 0.02$ but with different forcing functions. Panels (a) and (b)
        correspond to Taylor-Green forcing: (a) $u_y(y,z)$, and (b)
        $\theta(y,z)$. Panels (c) and (d) correspond to isotropic
        forcing: (c) $u_y(y,z)$, and (d) $\theta(y,z)$. Note the more
        evident horizontal winds in the simulation with isotropic
        forcing, and the different nature of the fluctuations in the
        temperature in both flows.}
    \label{cuts}
\end{figure*}

To ease with the understanding of the behavior caused by the two
different forcings, and to illustrate the presence of VSHW in the flows, 
in Fig.~\ref{cuts} we present vertical slices (at constant $x$) of the
potential temperature and horizontal velocity fields for the two 
simulations. Although the simulations have the same $\textrm{Fr}$ 
and $\textrm{Re}$, the flows are significantly different. On the one
hand, there are smaller scale structures and overturning in both
fields for the Taylor-Green forced simulation. On the other hand, 
the simulation with isotropic forcing shows a more dominant 
large-scale horizontal flow, and smaller scale (although smoother) 
fluctuations in the temperature. This is consistent with the fact that 
when forcing isotropically there is more energy in the waves, as seen 
in Fig.~\ref{tgn}. Another way to quantify how much energy is in the 
VSHW is to compare the ratio of the energy in the modes with 
$\omega=0$ (i.e., in the modes with $k_\perp=0$) to the total energy. 
For $\textrm{Fr}=0.02$, the Taylor-Green simulation has 
$\approx 0.009\%$ of its total energy in those modes, while the 
isotropically forced one has $\approx 0.03\%$, indicating that the 
isotropically forced simulation has stronger VSHW. Note however that
both simulations have a broad energy spectrum, indicating that
turbulence in both flows may be of different nature, with the case
with more energy in wave motions being smoother and possibly closer to
a wave turbulence regime.

The comparison between the two different forcing functions further indicates
that stratified turbulent flows may display different behavior depending on
whether the mechanism used to excite the turbulence allows or prevents the
development of large-scale vertically sheared horizontal winds. Interestingly,
the case studied here that injects more energy directly into the waves, is also
the case in which horizontal winds and CL instability more clearly develop, two
features that are not considered in wave turbulence theories.

\section{Conclusions} 

In a turbulent flow, waves and individual instability events cannot 
be easily identified. As a result, previous numerical and observational 
studies of Doppler shift and CL instability focused on analyzing
single wave packets travelling through a background flow. Our
analysis, based on computation of a four-dimensional spectrum with 
high temporal and spatial resolution, allowed us to study these 
phenomena in turbulent flows, and to identify direct evidence of 
their occurrence.

With these tools we showed that Doppler shift and CL instability 
occur naturally in a stratified disordered flow, as a result of the
interaction of the waves with the horizontal winds. This indicates 
the CL instability can be one of the mechanism behind the formation 
of large scales structures in stratified flows, often observed in
simulations but whose origin is unclear 
\cite{Smith02,Marino13,Marino14}. Moreover, although Doppler shift 
is observed in all forcing functions considered, development of the 
CL instability requires the external forcing not to disrupt the 
development of mean horizontal winds. Theories of stratified wave 
turbulence should take these effects into account. The mechanism, 
and the tools presented here, can be also relevant in 
quasi-geostrophic turbulence \cite{James87} and plasma turbulence 
\cite{Diamond05,Connaughton11}, where zonal flows are also known 
to develop.

\begin{acknowledgments}
The authors acknowledge support from Grant Nos. PIP 11220090100825,
UBACYT 20020110200359, PICT 2011-1529 and PICT 2011-1626. PCdL
acknowledges useful comments from the organizers and attendants of the
2014 ICTP Hands-On Research School.
\end{acknowledgments} 

\bibliography{ms}

\end{document}